# Thermoelectric signature of quantum critical phase in a doped spin liquid candidate

K. Wakamatsu[1], Y. Suzuki[1], T. Fujii[2], K. Miyagawa[1], H. Taniguchi[3] and K. Kanoda[1]*

[1]*Department of Applied Physics, University of Tokyo; Bunkyo-ku, Tokyo, 113-8656, Japan.*

[2]*Cryogenic Research Center, University of Tokyo; Bunkyo-ku, Tokyo, 113-0032, Japan.*

[3]*Graduate School of Science and Engineering, Saitama University; Saitama 338-8570, Japan.*

*Corresponding author. Email: kanoda@ap.t.u-tokyo.ac.jp

**Quantum spin liquid is a nontrivial magnetic state of longstanding interest, in which spins are strongly correlated and entangled but do not order[1, 2]; further intriguing is its doped version, which possibly hosts strange metal and unconventional superconductivity[3]. Promising and currently the only candidate of the doped spin liquid is a triangular-lattice organic conductor, κ-(BEDT-TTF)$_4$Hg$_{2.89}$Br$_8$, recently found to hold metallicity, spin-liquid-like magnetism and BEC-like superconductivity[4-6]. The nature of the metallic state with the spin-liquid behaviour is awaiting to be further clarified. Here, we report the thermoelectric signature that mobile holes in the spin liquid background is in a quantum critical state and it pertains to the BEC-like superconductivity. The Seebeck coefficient divided by temperature, *S*/*T*, is enhanced on cooling with logarithmic divergence indicative of quantum criticality. Furthermore, the logarithmic enhancement is correlated with the superconducting transition temperature under pressure variation, and the temperature and magnetic field profile of *S*/*T* upon the superconducting transition change with pressure in a consistent way with the previously suggested BEC-BCS crossover. The present results reveal that the quantum criticality in a doped spin liquid emerges in a *phase*, not at a point, and is involved in the unconventional BEC-like nature.**

Strong correlation among electrons brings about various emergent phenomena in solids. Among them, quantum criticality has long been a focus of profound interest since strange metal, unconventional superconductivity and magnetic quantum phase transition all spring from a single point called a quantum critical point as extensively discussed in heavy electron systems and copper oxides[7, 8]. Notably, a recent study of a heavy electron compound with a Kagome lattice, CePdAl, have found a quantum critical phase residing in a range of parameter space, not at a point, suggesting that a non-Fermi liquid phase is stabilised along with a quantum spin liquid (QSL) of *f* electrons[9]. The relation between the quantum critical phase and frustration has recently attracted intense attention[10, 11]. In this connection, it is notable that the organic conductor, κ-$(BEDT\text{-}TTF)_4Hg_{2.89}Br_8$ (abbreviated as κ-HgBr) is suggested to be a doped QSL that hosts a non-Fermi liquid phase in a finite pressure range.

κ-HgBr is a layered compound consisting of conducting BEDT-TTF layers with a nearly isotropic triangular lattice of BEDT-TTF dimers with the transfer integral ratio, $t'/t$, of 1.02 (Fig. 1b) and insulating $Hg_{2.89}Br_8$ layers. The nonstoichiometry of Hg comes from an incommensurate lattice against a BEDT-TTF lattice and the missing content from 3.0 contributes 11% hole doping to a half-filled band[12]. Remarkably, κ-HgBr shows non-Fermi liquidity and spin susceptibility well scaled to that of the spin liquid material, κ-$(BEDT\text{-}TTF)_2Cu_2(CN)_3$, thus suggesting that κ-HgBr hosts a doped QSL[5]. The electronic nature of κ-HgBr is indicated to alter by pressure. The Hall coefficient behaves such that charge carriers are apparently only the doped holes at low pressures due to strong correlation prohibiting double occupancy but is recovered to full band carriers at high pressures[4, 13]. This appears to be a pressure equivalence of the doping-driven $p$ to $1+p$ crossover with $p$ a doping content in cuprates[14]. In resistivity, the non-Fermi liquid persists up to around 0.4-0.5 GPa and crosses over or transitions to a Fermi liquid[4, 6, 15], as theoretically suggested[16]. At low temperatures, superconductivity occurs whose transition temperature, $T_c$, shows dome-like pressure dependence and whose nature changes from BEC-like to BCS condensate[6]. The schematic phase diagram is shown in Fig.1c.

The quantum criticality in a *phase* instead of at a point and its possible relevance to QSL is an issue of profound significance. The present work aims to verify the quantum critical nature of the electronic state in κ-HgBr with pressure variation through thermoelectric effect which is very susceptible to quantum criticality, exploiting highly compressible feature of organic crystal[17]. Here, we report our observation of the thermoelectric signature of quantum criticality in the doped QSL phase and its possible relevance to superconductivity.

Figure 2a and 2b shows the temperature dependence of the Seebeck coefficients divided by temperature, -*S*/*T*, under several pressures. Two separate measurements on different κ-HgBr samples give nearly coinciding results. To view the overall profile of -*S*/*T* in the pressure-temperature plane, we display the values with a range of colours in Fig. 2c for sample #1 (for sample #2, see Fig. S1 in Supplementary Information). -*S*/*T* behaves similarly at every pressure at high temperature above 30-40 K but, below that, shows strong pressure dependence with fan-shaped dispersion; -*S*/*T* is highly enhanced at low pressures well below 1 GPa whereas it is progressively reduced with increasing pressure. Sudden decreases in -*S*/*T* at low temperatures is due to the superconducting transition, as described in detail later.

At high pressures above 1 GPa, where the electron correlation is weakened, -*S*/*T* is constant at low temperatures, being consistent with the Fermi liquid behaviour observed in resistivity[4, 6, 15] (see also Fig. S2 in Supplementary Information). For Fermi liquids, *S* is expected to follow the formula[18],

$$\frac{S}{T} = \pm\frac{\pi^2}{3}(1+\lambda)\frac{k_B}{e}\frac{1}{T_F} \ , \quad (1)$$

where $T_F$ is the Fermi temperature and $\lambda$ is a parameter related to the energy dependence of relaxation time, e.g. $\lambda$=0 in the case of constant (energy-independent) mean free path. The -*S*/*T* values at low temperatures are 0.4-0.53 $\mu V/K^2$ at 1.50-1.55 GPa, ~0.75 $\mu V/K^2$ at 1.2 GPa, and ~1.1 $\mu V/K^2$ at 1.0 GPa and, assuming $\lambda$=0, these -*S*/*T* values yield $T_F$ =530-710, ~380 and ~260 K at 1.50-1.55, 1.2 and 1.0 GPa, respectively. $T_F$ decreases with pressure toward zero around 0.5 GPa (Fig. 3a), which is very probably ascribable to the progressive renormalisation of the Coulomb interaction. Concomitantly, the temperature dependence of -*S*/*T* starts to deviate from the Fermi liquid behaviour of *S/T* = constant.

With further decreasing pressure below 1 GPa, the temperature dependence of -*S*/*T* deviates appreciably from the Fermi liquid behaviour and -*S*/*T* continues to increase on cooling until superconductivity sets in at $T_c$. The low-temperature value just above $T_c$ reaches the values over 2.0 $\mu V/K^2$ at 0.5-0.65 GPa and levels off at lower pressures (Fig. 3a), where non-Fermi liquid behaviour of resistivity is observed[4, 6, 15]. The temperature profile of -$S/T$ in the low-pressure region is roughly linear in Fig. 2a, meaning -$S/T \propto \ln T$. Such temperature dependence of the Seebeck coefficient is observed as a signature of quantum criticality in strongly correlated systems such as cuprates[19-21], iron pnictide[22, 23], heavy fermion[24-28] and cobalt oxides[29], and intensively studied theoretically[30-33]. To be quantitative, the temperature dependences of -$S/T$ under pressured below 0.5 GPa were fitted by the form of $S/T = \gamma' \ln(T/T_0)$, where $T_0$ is a parameter of the energy scale of quantum critical fluctuations[30]. The fitting yields $T_0$=50-60 K, which is compared to $T_0$~170K for Nd-LSCO and $T_0$~3 K for $YbRh_2Si_2$ (ref.[19, 24]); these values appear consistent with their relative sizes of bandwidths of organic conductors, cuprates and heavy electron systems.

Thus, the present observation provides evidence for quantum criticality in the low-pressure region in κ-HgBr. A distinctive feature from the conventional cases is that it is extended in a finite pressure range, namely, in a “critical region” instead of a “critical point”. In the present case, both the magnitude and logarithmic behaviour of -$S/T$ maintain unchanged below 0.5-0.65 GPa as seen in Figs. 2a and 2b. In conjunction with another material that is suggested to host such a phase[9], spin frustration would be a key to the stabilisation of quantum critical phase, as suggested theoretically[10, 11]. It is noted that the enhanced $S/T$ values are not sharply suppressed upon the crossover from the non-Fermi liquid to the Fermi liquid at around 0.5 GPa. We consider this as a possible manifestation of the strong electron correlation in the marginal Fermi liquid nearby a non-Fermi liquid. The coefficient $A$ in the temperature dependence of resistivity, $\rho = \rho_0 + AT^2$, in the Fermi-liquid regime is a measure of correlation strength or quasi-particle dumping rate. The pressure dependence of $A$ measured with a separate sample is displayed in the inset of Fig. 3a, which exhibits its remarkable increase well before entering the non-Fermi liquid regime.

In many cases, the quantum critical logarithmic-in-temperature evolution of -$S/T$ appears in the vicinity of magnetic transitions[30]. In κ-HgBr, enhanced spin fluctuations are suggested by NMR studies[34] and thus likely involved in the enhanced $S/T$ albeit in a different way from the magnetic quantum criticality because of no magnetic order in κ-HgBr. It is known that $S$ is empirically well expressed by $S$~$C/ne$ at temperatures, where $C$ is specific heat and $n$ is the density of charge carriers with charge $e$; thus, $S$ is roughly an entropy per charge carrier[35]. As indicated by the Hall coefficient and resistivity, the nature of charge carriers is changed with decreasing pressure from the band quasiparticles with Fermi liquidity to emergent holes (that should be called holons) with non-Fermi liquidity, which originate from the prohibition of double occupancy – akin to the Mott localisation in a half-filled system. This drastic change of the carrier nature with no magnetic symmetry breaking should affect the enhanced $S/T$. The emergent holons may have extraordinary charge fluctuations that are entangled with a QSL having large entropy[36]. The reported values of |γ'|, a possible measure of the strength of quantum fluctuations[30], are 0.01-0.05 $\mu V/K^2$ for electron-doped cuprates[21], 0.05-0.11 $\mu V/K^2$ for hole-doped cuprates[19, 20], 0.3-0.9 $\mu V/K^2$ for ion pnictides ($Ba(Fe_{1-x}Co_x)_2As_2$) (ref.[22]), and 2.3, 4.5, 6.2 $\mu V/K^2$ for heavy electron systems (UCoGe, $YbRh_2Si_2$, and $CeCu_{5.9}Au_{0.1}$, respectively)[24-26] (see Table S1 in Supplementary Information for |γ'| values of other materials). Thus, the present |γ'| value for κ-HgBr, ~1.2 $\mu V/K^2$, suggests a relatively large coupling of the quantum criticality in the doped QSL to the thermoelectric effect.

Remarkably, the superconducting transition temperature $T_c$ is well correlated with the low-temperature values of the logarithmically enhanced $-S/T$, as seen in Figs. 3a and 3b (see Fig. S3 and S4 in Supplementary Information for definitions of $T_c$ and onset $T^*$, and the pressure dependences of |γ'| and $T_c$, respectively). Given that |γ'| is an indicator of the strength of critical fluctuation[30], the correlation suggests that the critical fluctuations mediate or facilitate the electron pairing. Such correlation is also found in cuprates and iron pnictides as well[21, 22]. Figure 4 shows the low-temperature behaviour of $-S/T$ upon superconducting transition under zero and applied magnetic fields perpendicular to the conducting layers. $-S/T$ vanishes in the superconducting state. The transition is quite sharp at high pressures; at lower pressures, however, it becomes rounded with an onset well prior to the bulk transition (Fig. 3b), reserving the possibility of enhanced superconducting fluctuations at lower pressures. Figure 4 also shows that the superconductivity is entirely destroyed by a field of 3 T under 0.9 GPa whereas, under 0.3 GPa, it survives even at a field of 7 T albeit partially very probably as a vortex liquid state. These superconductive features are fully consistent with the previously revealed BEC-to-BCS crossover associated with a non-Fermi liquid to a Fermi liquid crossover in κ-HgBr (ref.[6]). In the BEC-like regime at low pressures, the superconductivity shows enhanced fluctuations with preformed Cooper pairs and is robust to magnetic field with forming a vortex liquid state[6]. Thus, the present results lend support to the picture of pressure-induced BEC-BCS crossover in κ-HgBr from thermoelectric point of view.

The present work is the first thermoelectric investigation of a doped spin liquid candidate. The thermoelectric effect probes the entropy transport by charge carriers, which are influenced by magnetic background and superconductivity if any, and therefore includes information on the surroundings of the doped holes. The logarithmic Seebeck enhancement observed at low pressures signifies that charge carriers that travel, avoiding double occupancies, in the sea of spin liquid suffer from quantum critical fluctuations in charge and/or spin degrees of freedom. It is emphasised that the quantum critical state resides as a phase, not at a point. As pressure is increased, the logarithmic enhancement is suppressed and crosses over to the conventional metallic behaviour, indicating that the doped spin liquid crosses over to a Fermi liquid by reducing the Coulomb interactions. The correlation between the logarithmic Seebeck enhancement and superconductivity suggests that the anomalous quantum critical fluctuations favor the BEC-like electron pairing. It is an issue of further investigation whether spin or charge fluctuations or both mediate the Cooper pairing.

**Acknowledgments.** We thank H. Oike for comments and fruitful discussion. This work was supported by Japan Society for the Promotion of Science (JSPS) under Grant Numbers 18H05225, 19H01846, 20K20894, 20KK0060 and 21K18144, and by JST SPRING under Grant Number JPMJSP2108. Most parts of this work were performed using facilities of the Cryogenic Research Center, the University of Tokyo.

**Author contributions.** K.K. designed the project. H.T. prepared samples. K.W., Y.S, T.F. and K.M. performed experiments and analysed as well as interpreted data with the help of K.K. K.W. and K.K. wrote the manuscript with the input from all authors.

**Competing financial interests.** The authors declare no competing financial interests.

**Additional information**

**Supplementary information is available for this paper at ***.**

**Correspondence and requests for materials should be addressed to K.K.**

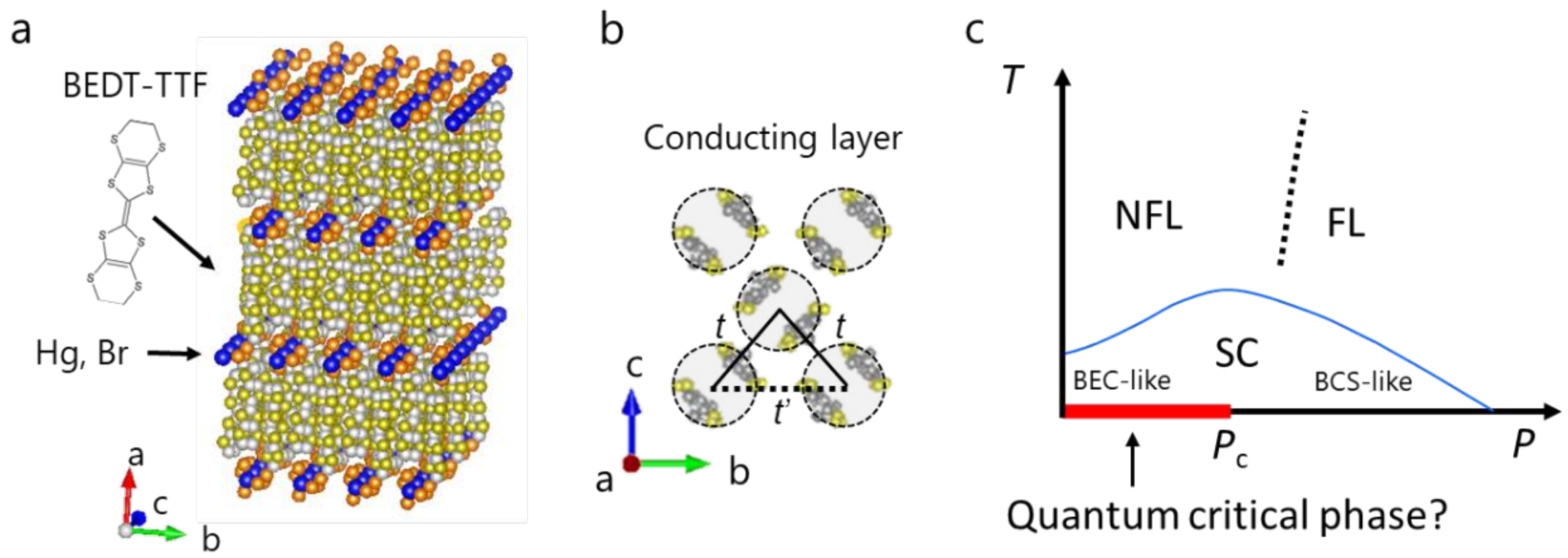

**Fig. 1 | Crystal structure and schematic phase diagram of κ-HgBr.**

**a**, Layered crystal structure of κ-HgBr. The orange and blue spheres indicate Br and Hg ions in the insulating layers, respectively. **b**, In-plane molecular arrangement in the conducting layer of κ-HgBr. The BEDT-TTF molecules form dimers (circled by dotted lines), which construct an isosceles triangular lattice, characterised by two kinds of transfer integrals of $t$ and $t'$ between the adjacent antibonding dimer orbitals; the ratio, $t'/t$, is 1.02 according to the molecular orbital calculations (see Supplementary Information). **c**, Schematic pressure-temperature phase diagram of κ-HgBr drawn with reference to the previous studies[4, 6, 15]. The NFL, FL, SC, and $P_c$ stand for non-Fermi liquid, Fermi liquid, superconductivity, and critical or crossover pressure between NFL and FL, respectively. The red bold line indicates the possible critical phase inferred from ref.[4, 6, 15].

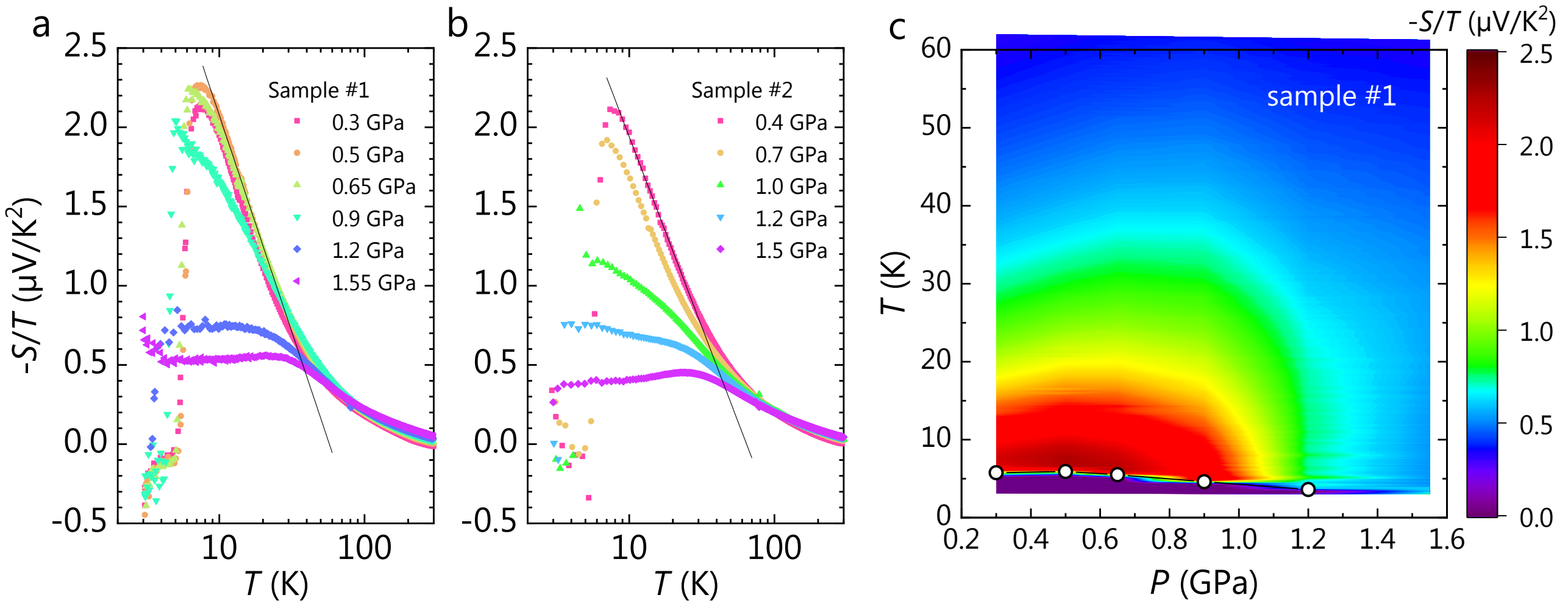

**Fig. 2 | Temperature and pressure profiles of -*S*/*T* in κ-HgBr.**

**a and b**, Temperature dependence of the Seebeck coefficient divided by temperature, $-S/T$, in the samples #1 and #2, respectively. The essential features of the results coincide with each other. Thin straight lines indicate the dependence of $-S/T \propto \ln T$. **c**, Contour plot of $-S/T$ in the pressure-temperature plane for the sample #1. The contour plot for the sample #2 is shown in Fig. S1 in Supplementary Information. The open circles indicate the superconducting transition temperature $T_c$ determined from

-$S/T$ (see Fig. 4 and Supplementary Information for definition of $T_c$).

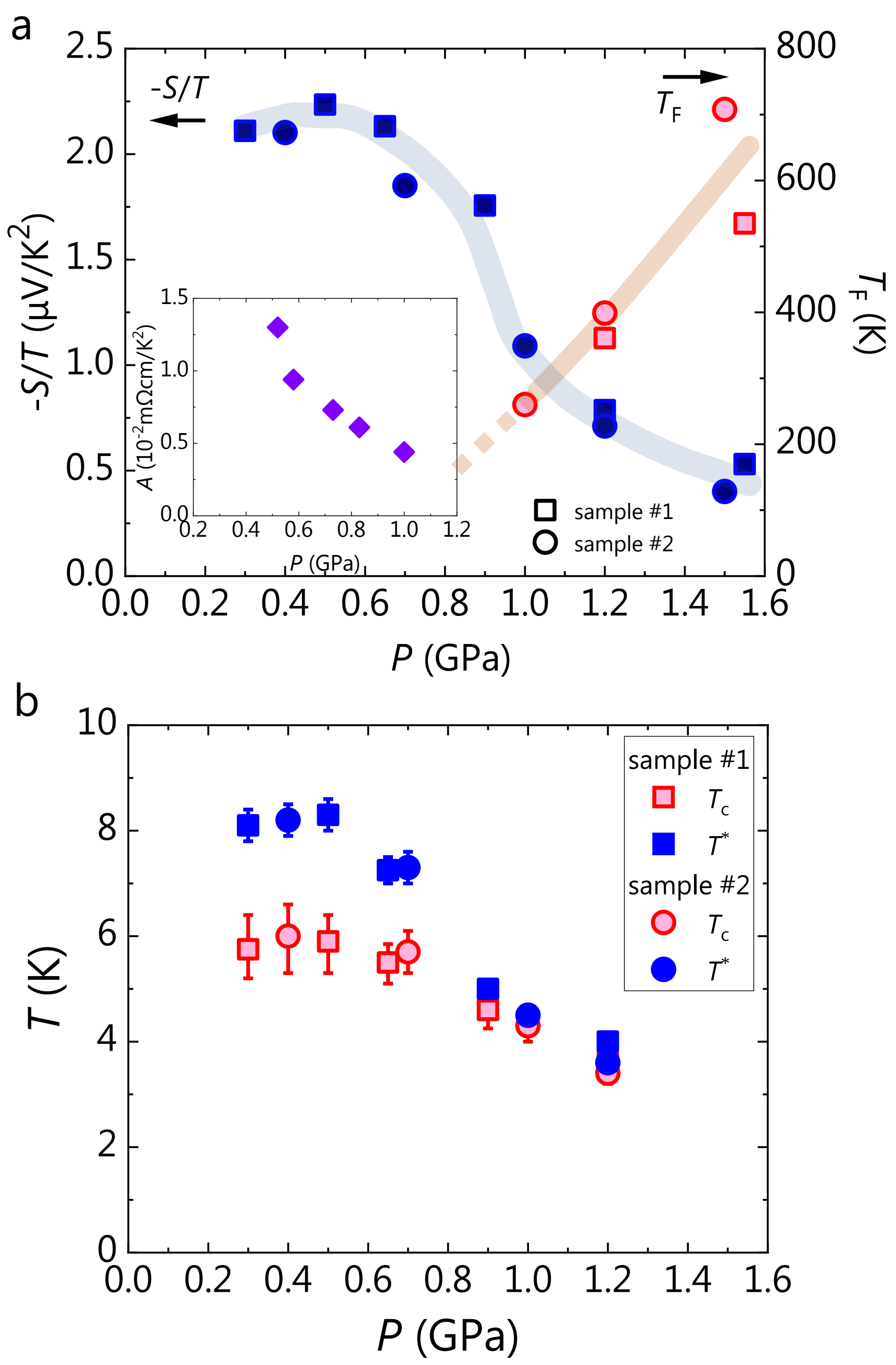

**Fig. 3 | Pressure dependences of -$S/T$ and superconducting transition temperature in κ-HgBr.** **a,** Pressure dependences of the -$S/T$ value at 8 K, just above $T_c$, and the Fermi temperature, $T_F$. The square and circle markers correspond to the samples #1 and #2, respectively. The blue and red markers are the -$S/T$ and $T_F$ values, respectively. Inset shows the pressure dependence of the coefficient, $A$, in the fit of the form, $\rho=\rho_0+AT^2$, to the separately measured resistivity. **b**, Pressure dependences of the superconducting transition temperature, $T_c$, and its onset, $T^*$. The error bars of $T_c$ indicate the widths of the bulk superconducting transition. $T^*$ is defined as the temperature at which -$S/T$ starts to deviate from the normal-state behaviour. The definitions of $T_c$, its error bar and $T^*$ are described in Supplementary Information.

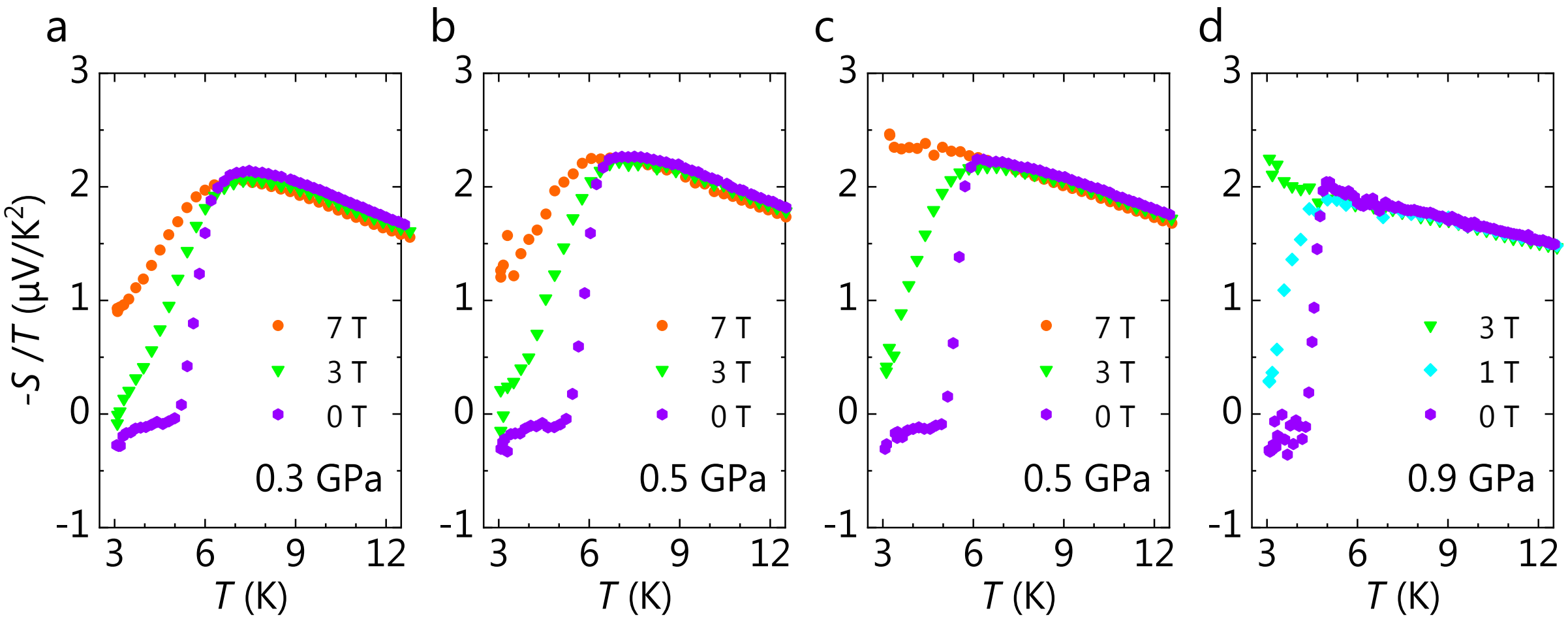

**Fig. 4 | Temperature dependence of -*S*/*T* at zero and applied magnetic fields.**

**a-d**, Temperature dependence of -*S*/*T* (for the sample #1) under the pressures of 0.3 (a), 0.5 (b), 0.65 (c) and 0.9 GPa (d). Magnetic field was applied perpendicular to the conducting plane.

## Methods

Single crystals of κ-HgBr were grown in the standard electrochemical method. For pressurisation, we used a clamp-type piston-cylinder cell made of CuBe/NiCrAl and Daphne oil 7373 as a pressure-transmitting media. Daphne oil 7373 solidifies on cooling so that the clumped pressure gradually decreases by 0.15-0.2 GPa as the temperature decreases from 300 K to 50 K and then takes a nearly constant value at lower temperatures. To know the internal pressure in the piston-cylinder cell, we used $T_c$ of a Sn flake that was mounted in the cell. The pressure values quoted in this article are the internal pressures thus estimated.

Thermoelectric effect was measured with a conventional experimental platform where two Cu-plates with the Cernox thermometers attached on both and a heater attached on one plate are bridged by a κ-HgBr crystal. The thermometers were calibrated at each pressure using a reference thermometer. The heater generated a temperature difference, $\Delta T$, between the two Cu-plates, which was maintained less than $T/10$ throughout the experiments. With measuring thermoelectric potential difference, $\Delta V$, between the plates under the temperature deference, $\Delta T$, the Seebeck coefficient is defined by $S=\Delta V/\Delta T$. In the present experiment, temperature gradient was applied along the c-axis in the conduction plane (Fig 1b).

The rapid cooling is often detrimental to organic conductors because it may cause crystal cracking and/or conformational disorder of terminal ethylene groups in BEDT-TTF. To minimise these possible faults, we cooled the sample at rates slower than 0.5 K/min. Even with such cautious cooling process, the cracking in κ-HgBr crystal was not avoided at ambient pressure. Thus, the present experiments were performed under pressures, where the sample was free from such a problem.

Supplementary Information for

# Thermoelectric signature of quantum critical phase in a doped spin liquid candidate

## Crystal structure and band filling

Most of κ-type BEDT-TTF compounds have the composition of κ-(BEDT-TTF)$_2$*X* with anion, *X*. κ-(BEDT-TTF)$_2$*X* has a quasi-two-dimensional structure with conducting BEDT-TTF layers and insulating *X* layers alternately stacked. In the conducting layer, BEDT-TTF dimers form an isosceles triangular lattice characterised by two different transfer integrals, *t* and *t*'. In the insulating layer, anion, *X*, electronically has a closed shell structure with valence of -1. Therefore, the valence of BEDT-TTF is +0.5; namely, one hole resides per one dimer site so that the band is half-filled in most κ-(BEDT-TTF)$_2$*X* systems.

κ-(BEDT-TTF)$_4$Hg$_{2.89}$Br$_8$ is exceptional in that the lattice periodicity of Hg ions in the insulating layer is incommensurate with that of BEDT-TTF molecules in conducting layers[1,2], resulting in a deviation from the half filling of the band[3]. The incommensurability, namely, the nonstoichiometry in κ-(BEDT-TTF)$_4$Hg$_{2.89}$Br$_8$ is not varied but fixed by chemistry to give 11% hole doping to half-filing. Note that mercury-included compounds often have incommensurate structures[4].

The ratio, *t*'/*t*, in κ-HgBr is 1.02 according to Ref.[5], where the *t* and *t*' values are evaluated by tight-biding approximations based on the extended Hückel calculations of molecular orbitals.

## Contour plot of -*S*/*T* of sample #2

Figure S1 shows the contour plot of -*S*/*T* of sample #2 in pressure-temperature plane. The open circles are superconducting transition temperature evaluated from the temperature dependence of -*S*/*T*. The contour plot of sample #2 well reproduces that of sample #1 shown as Fig. 2b in the main text.

## Temperature dependences of the resistivity and -*S*/*T* at 0.4 and 1.5 GPa

Figures S2a and S2b show the temperature dependences of -$S/T$ and the resistivity of sample #2. The resistivity is normalised to the value at 15 K. The Seebeck coefficient and resistivity were measured simultaneously at the same setup. At 1.5 GPa, the resistivity shows the squared temperature dependence and -$S/T$ is nearly constant at low temperatures, confirming the Fermi liquid behaviour. At 0.4 GPa, the resistivity shows linear-in-temperature dependence down to $T_c$ and -$S/T$ exhibit the logarithmic divergence down to $T_c$.

**Definition of $T_c$ and $T^*$**

Figure S3 shows the temperature dependence of -$S/T$ at zero field under 0.3 GPa for sample #1. The uppermost and lowermost values and the midpoint defined in Fig. 3b correspond to the upper and lower bounds of the error bar. The onset temperature, $T^*$, is defined as the temperature at which -$S/T$ begins to deviate from the high-temperature normal-state behaviour.

**Pressure dependence of |γ'| and $T_c$**

Figure S4 shows the pressure dependences of |γ'| and $T_c$. The γ' value was evaluated by fitting the form, $S/T = \gamma'\ln(T/T_0)$, to the experimental data. The $T_c$ is the midpoint as described above.

**γ' values of $S/T = \gamma'\ln(T/T_0)$ for other materials**

Table S1 lists the |γ'| values reported in literatures[6-18], where γ' values were evaluated from the behaviour of $S/T = \gamma'\ln(T/T_0)$.

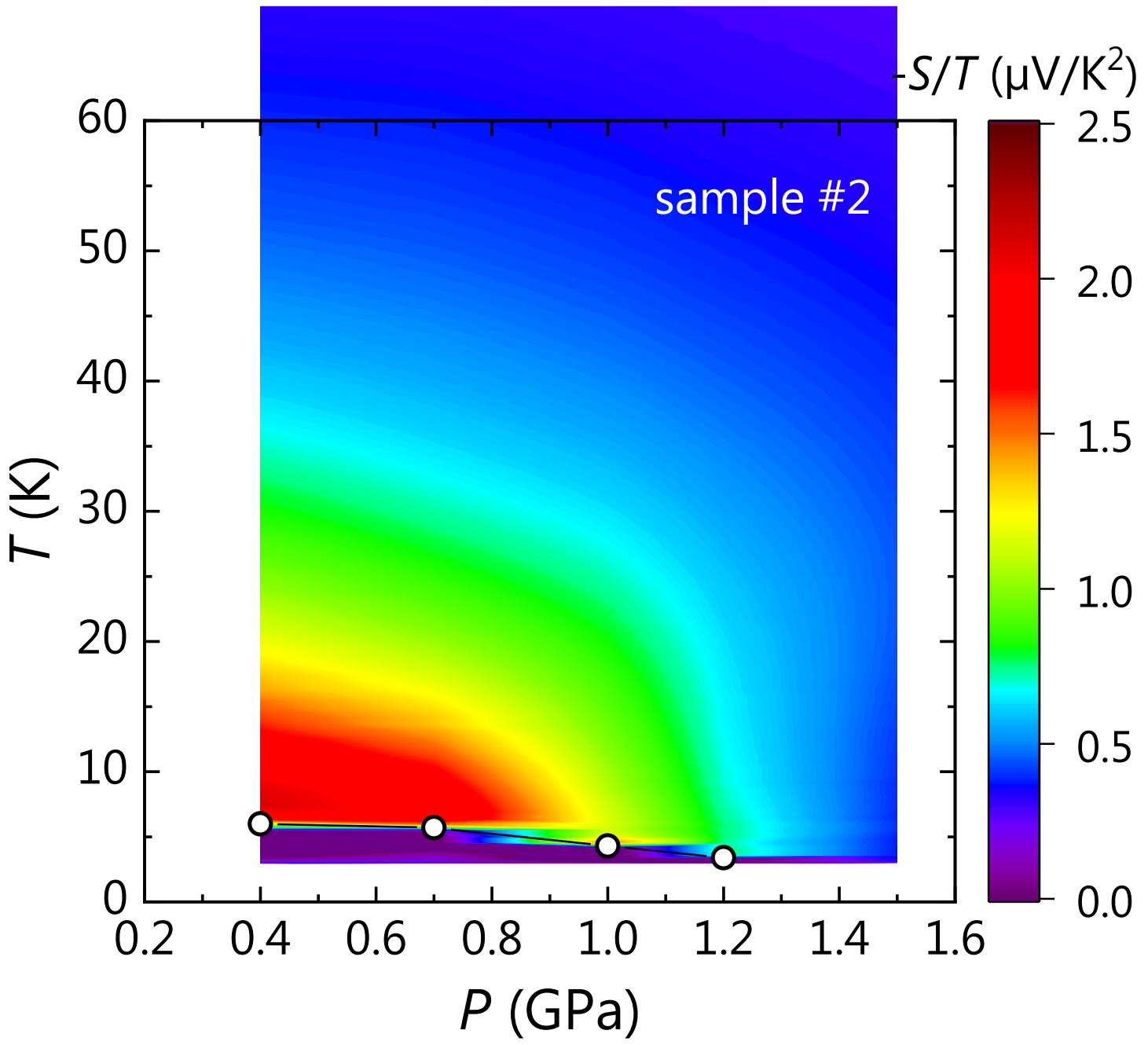

**Figure S1 | Contour plot of -*S*/*T* in κ-HgBr for sample #2.**

The graph shows the contour plot of -*S*/*T* for sample #2 in the pressure-temperature plane. The open circles indicate the superconducting transition temperature.

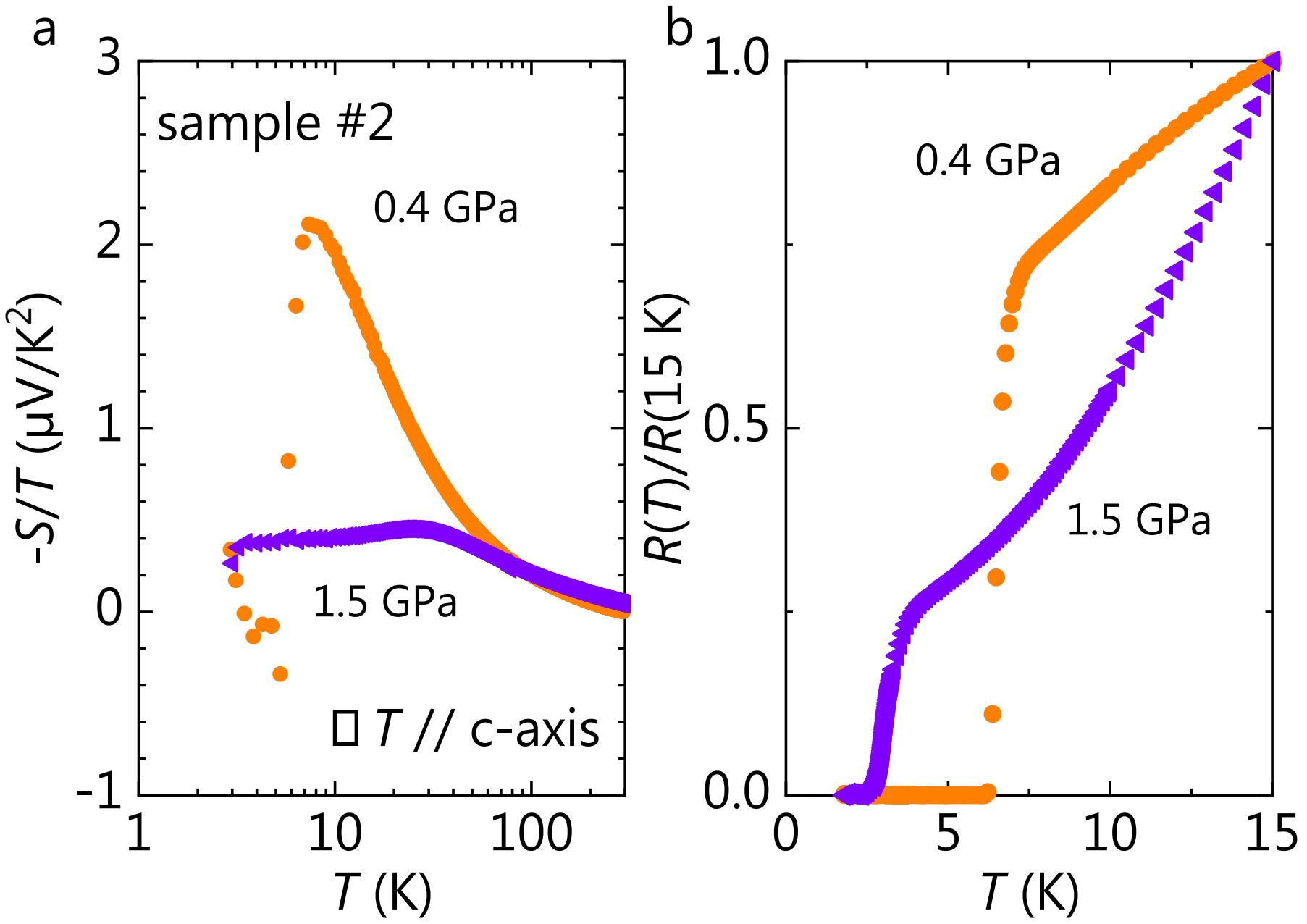

**Figure S2 | Temperature dependences of the -*S*/*T* and resistivity simultaneously measured for sample #2 at 0.4 and 1.5 GPa.**

**a**, Temperature dependence of -*S*/*T* in the logarithmic temperature scale. At 0.4 GPa, -*S*/*T* exhibits ln*T* behaviour. At 1.5 GPa, however, the Fermi liquid behaviour of -*S*/*T*=constant was observed. **b**, Temperature dependence of the resistivity normalised to its value of 15 K. The linear- and quadratic-in-temperature resistivity above $T_c$ under 0.4 and 1.5 GPa indicate non-Fermi liquid and Fermi liquid behaviours, respectively.

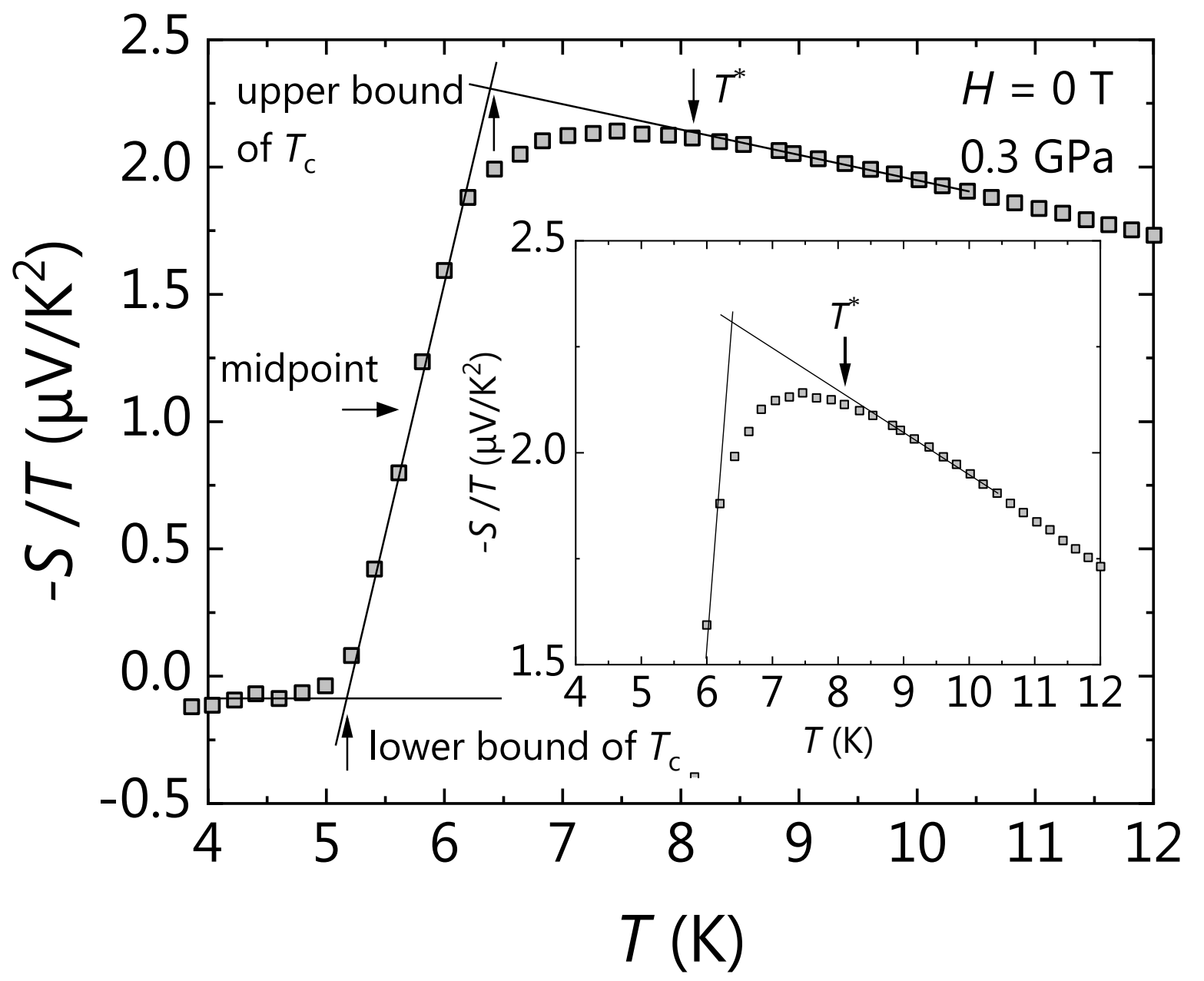

**Figure S3 | Definition of $T_c$ and $T^*$.**

The definitions of bulk $T_c$, its error and the onset, $T$*, are shown for the 0.3 GPa data of -$S$/$T$ of sample #1, for example. The bulk $T_c$ is defined by the midpoint. The upper and lower bounds of $T_c$ indicated in the figure determines the error bar of $T_c$. $T^*$ is defined as the temperature at which -$S$/$T$ starts to deviate from the normal-state behaviour. The inset shows the enlarged view of the -$S$/$T$ behaviour in the vicinity of $T^*$.

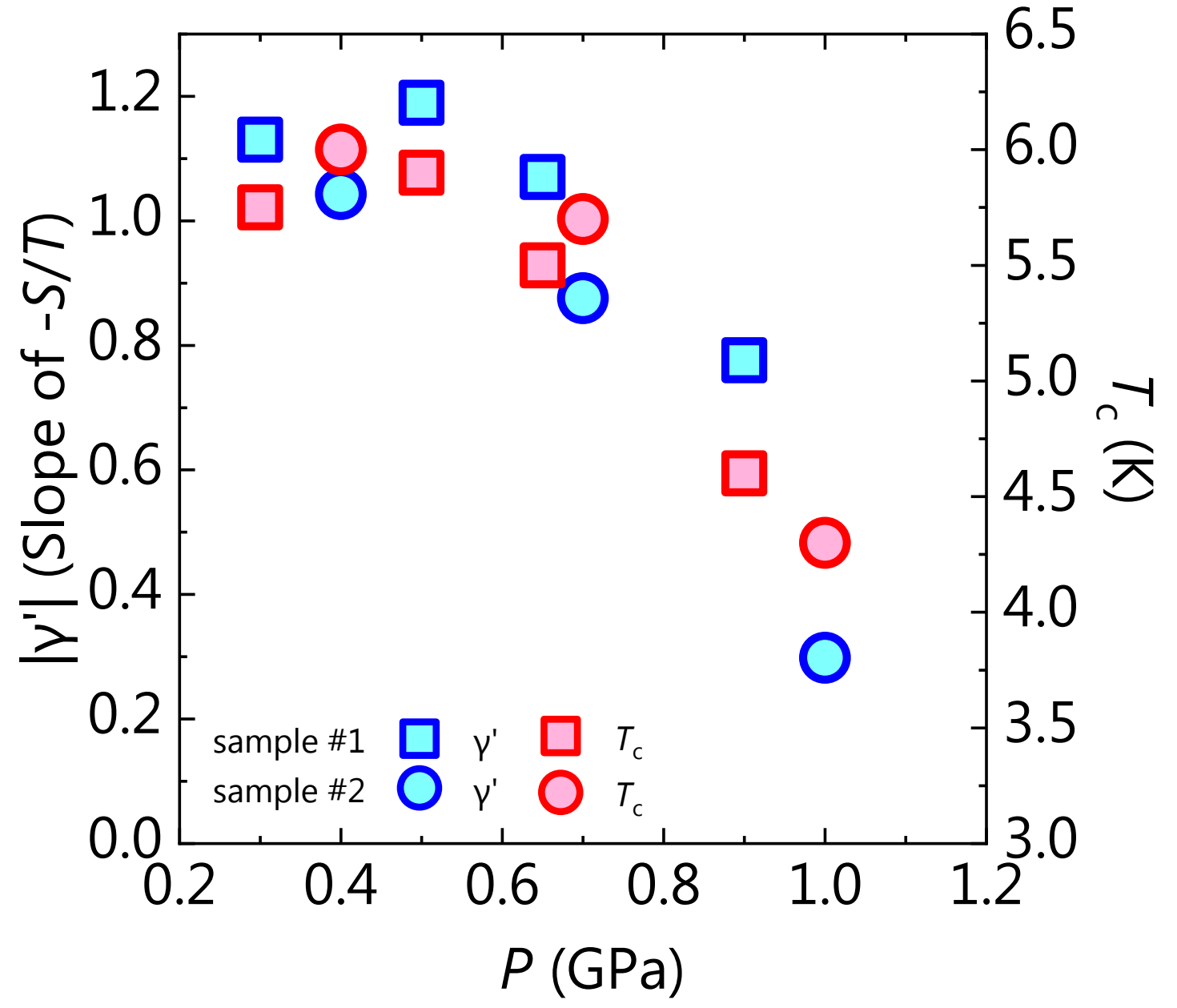

**Figure S4 | Pressure dependence of |γ'| and $T_c$ in κ-HgBr.**

The |γ'| and $T_c$ values for sample #1 and #2 are plotted as a function of pressure. The squares and circles correspond to the data of sample #1 and #2, respectively. The blue and red symbols indicate |γ'| and $T_c$, respectively.

**Table S1 | |γ'| values in the logarithmic part of -*S*/*T* in some materials**

The |γ'| values in the behaviour of $S/T = \gamma' \ln(T/T_0)$ for other materials[6-18] are listed in $\mu V/K^2$.

| Material | Slope value | Material | Slope value |
|---|---|---|---|
| Nd-LSCO ($p$ = 0.24) (ref.[6]) | 0.11 | UCoGe (H=11.1 T) (ref.[12]) | 2.3 |
| Eu-LSCO ($p$ = 0.24) (ref.[7]) | 0.16 | $YbRh_2Si_2$ (ref.[13]) | 4.5 |
| Bi2201 ($p$ = 0.39) (ref.[8]) | 0.05 | $CeCu_{5.9}Au_{0.1}$ (ref.[14]) | 6.2 |
| PCCO ($x$ = 0.16-0.19) (ref.[9]) | 0.012-0.038 | $Ce_2PdIn_8$ (ref.[15]) | 1.6 |
| LCCO ($x$ = 0.15-0.17) (ref.[9]) | 0.0095 – 0.049 | YbAgGe (ref.[16]) | 4.7 |
| $[BiBa_{0.66}K_{0.36}O_2]CoO_2$ (ref.[10]) | 0.62 | YbPtBi (ref.[17]) | 0.25 |
| $Ba(Fe_{1-x}Co_x)_2As_2$ (p=0.022-0.13) (ref.[11]) | 0.32 - 0.895 | $EuFe_2(As_{1-y}P_y)_2$ (y=0.26, 0.36) (ref.[18]) | 0.077, 0.16 |